\documentclass{elsart}

\usepackage{floatflt,graphicx,hyperref}
\usepackage{amssymb}

\begin{document}

\begin{frontmatter}

% Title, authors and addresses

% use the thanksref command within \title, \author or \address for footnotes;
% use the corauthref command within \author for corresponding author footnotes;
% use the ead command for the email address,
% and the form \ead[url] for the home page:
% \title{Title\thanksref{label1}}
% \thanks[label1]{}
% \author{Name\corauthref{cor1}\thanksref{label2}}
% \ead{email address}
% \ead[url]{home page}
% \thanks[label2]{}
% \corauth[cor1]{}
% \address{Address\thanksref{label3}}
% \thanks[label3]{}

\title{The Radio Cerenkov Technique for Ultra-High Energy Neutrino Detection}
%\title{Detection of Cerenkov Radiation of Very Long Wavelength (Radio Region)}

% use optional labels to link authors explicitly to addresses:
% \author[label1,label2]{}
% \address[label1]{}
% \address[label2]{}

\author{Amy Connolly}

\address{Department of Physics and Astronomy, University College London \\
Gower Street \\
LONDON \\
WC1E 6BT 
}
\begin{abstract}
I review the status of the Radio Cerenkov detection technique
in searches for ultra-high energy (UHE) neutrinos of cosmic origin.
After outlining the physics motivations for UHE neutrino searches, I
give an overview of the status of 
current and proposed experiments in the field.

% Text of abstract
\end{abstract}

\end{frontmatter}

% main text
\section{Introduction}
\label{intro}
Over the past century,
cosmic radiation from sources outside the earth's atmosphere 
has been measured over 
14 orders of magnitude in energy.   
At $10^{18}$~eV, cosmic rays are incident on the earth's atmosphere 
at a rate of 
one particle/km$^{2}$/year.
Cosmic ray experiments HiRes and Auger~\cite{hiresauger} both
see a break in the spectrum near $10^{19.5}$~eV and this was expected;
at this energy, protons and
heavy nuclei reach their threshold for interacting with cosmic microwave 
background (CMB)
photons through what is known as the Greisen-Zatsepin-Kuzmin (GZK) 
process~\cite{gzk}.
This process can, for example, occur through a $\Delta$ resonance:
\begin{center}
\begin{equation}
\label{eq:gzk}
p + \gamma_{\mathrm{CMB}} \rightarrow \Delta^{*} \rightarrow n ~\pi^{+} \rightarrow n ~e^{+} ~\nu_{\mu} ~\nu_{e} ~\bar{\nu_{\mu}}
\end{equation}
\end{center}
The GZK process slows a cosmic ray above threshold within 
approximately 50~Mpc of its source.
Above a few TeV, gamma rays too are absorbed
before reaching us, but in the infrared background
radiation through pair production~\cite{gammaraycutoff}.

\begin{figure}[htb]
\begin{minipage}{0.47\textwidth}
  \begin{center}
  \scalebox{0.37}{\includegraphics{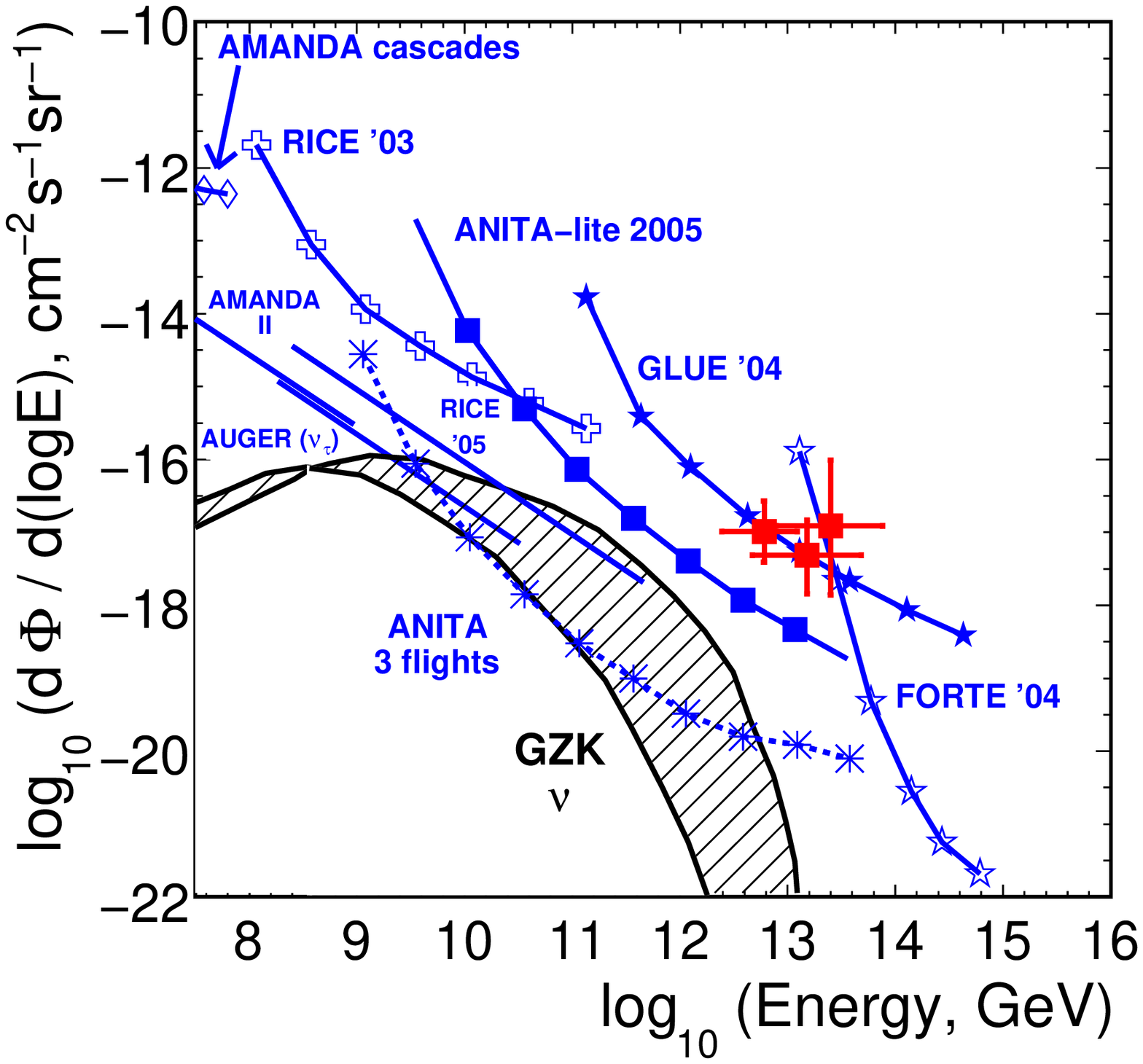}}
\caption{From~\cite{anitalite} with recent 90\%~CL 
limits from RICE `05, Amanda II and
AUGER added~\cite{rice05,amandaiiicecube,augerneutrinos}.  The ANITA curve is
an expected limit established pre-flight.  The shaded region is
a range of expected neutrino fluxes from the GZK process.}
\label{fig:neutrinos}
\end{center}
\end{minipage}
\hfill
\begin{minipage}{0.47\textwidth}
  \begin{center}
\scalebox{0.33}{\includegraphics{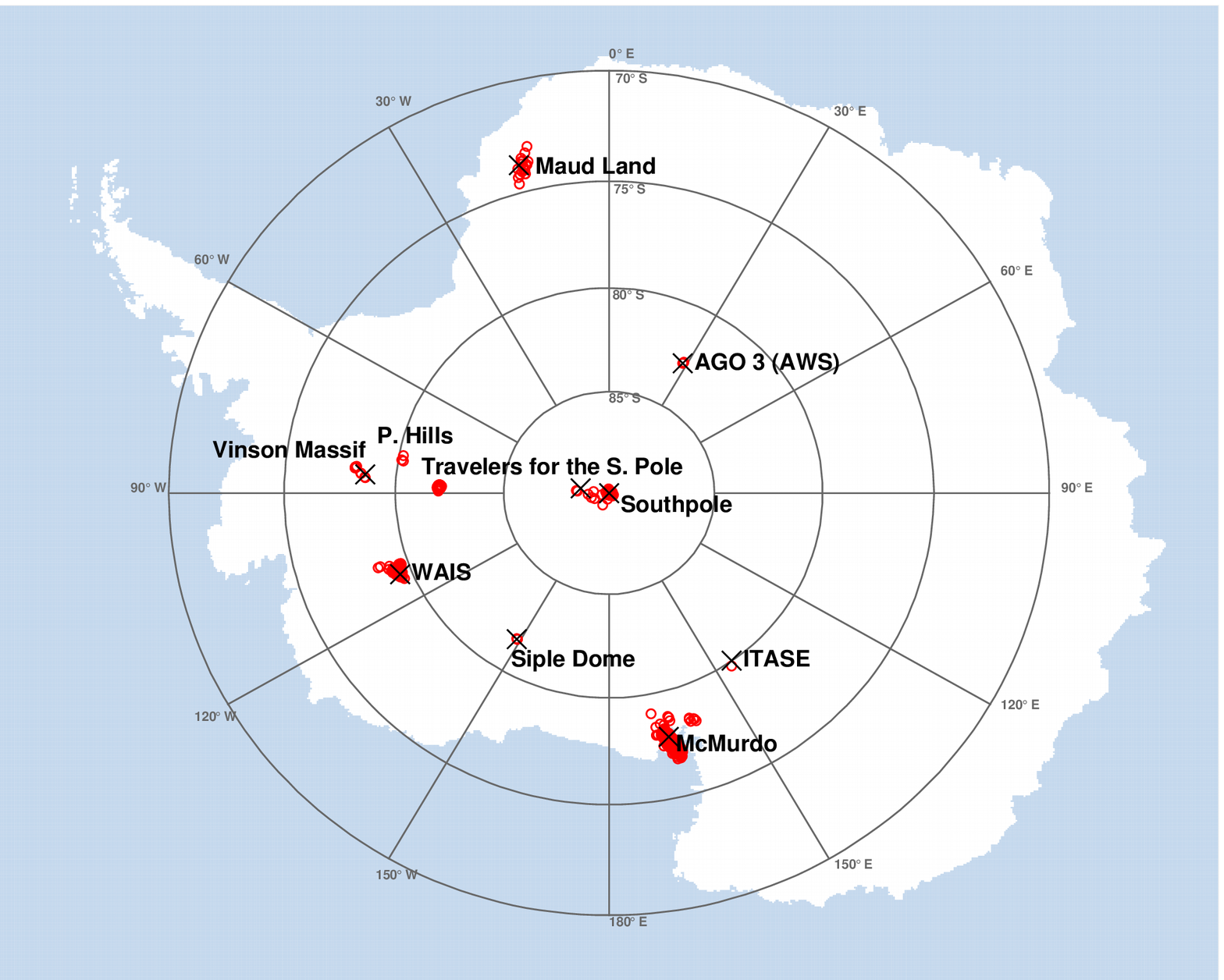}}
\vspace{0.35in} 
  \caption{Map of reconstructed source locations from 10\% of the
data from the ANITA flight.  All events
in this 10\% data set have been
associated with camps, travelers and automatic weather stations. 
Plot by Jiwoo Nam.}
  \label{fig:map}
  \end{center}
\end{minipage}
\end{figure}

While the GZK process prevents us from observing distant cosmic rays above
the cutoff, the same process is a source of neutrinos, 
as seen in Equation~\ref{eq:gzk}, 
which can travel cosmic distances unattenuated.  
In addition, any photohadronic process that produces cosmic rays
within an astronomical source would be expected to produce neutrinos
from the decay of charged pions.  
Figure~\ref{fig:neutrinos}
includes a range of models for the expected neutrino flux from the
GZK process in the UHE 
region and current limits on diffuse neutrino fluxes.

To date, the only observed sources of neutrinos outside of
the earth's atmosphere have been the sun~\cite{sun} and 
Supernova~1987a~\cite{supernova1987a}.  
No diffuse cosmic neutrino flux has been measured, though
such a measurement is crucial to a complete picture of the UHE universe.  
Neutrinos are the 
only messengers above $10^{19.5}$~eV that can probe
distances greater than hundreds of Mpc.  
Undeflected by magnetic
fields, neutrinos point back to their place of origin,
and above $10^{17}$~GeV
would produce interactions with center-of-mass energies exceeding that of
typical
interactions at the LHC.

Several experiments are searching for a cosmic flux of neutrinos 
by detecting the visible Cerenkov radiation produced by
particle tracks~\cite{otherexperiments},
the largest being
the AMANDA-II experiment and its km$^3$ successor 
IceCube~\cite{amandaiiicecube}.  Their observations are
consistent with the expectation from atmospheric neutrino 
background.  They constrain the neutrino flux
below $10^{18}$~eV.

\section{Radio Cerenkov}
\label{cerenkov}

At and above $10^{18}$~eV, the expected neutrino flux 
is too small to be detectable in the km$^{3}$-size
volumes that are characteristic of the visible Cerenkov technique.  
Although theoretical predictions for neutrino fluxes vary by 
orders of magnitude, a mid-range model gives approximately 
10 neutrinos/km$^{2}$/year.  If a $10^{18}$~eV neutrino interacts
in ice within 300~km, we expect 0.03 neutrino
interactions/km$^{3}$/year.  
Once we account for a typical detector acceptance,
the rate becomes
one interaction in a km$^{3}$ detection volume every 100 years.

Radio Cerenkov is the most established neutrino detection 
technique that has
the scalability to sustain an UHE 
neutrino particle and astrophysics program over the coming decade.
The radio Cerenkov signal was predicted by Gurgen Askaryan in 
1962~\cite{askaryan}.
The signal originates from the
 net current in a shower rather than from individual tracks.
As an electromagnetic shower develops,  
a 20\% charge asymmetry 
appears as electrons
in the medium become part of the shower, and likewise positrons 
in the shower are annihilated by electrons in the medium.  
Since the transverse size of the shower is of order 10~cm, the
Cerenkov radiation is coherent only for wavelengths longer than
that (frequencies $<\mathcal{O}\left(\mathrm{1~GHz}\right)$).
This effect has been confirmed 
experimentally in accelerator beam tests~\cite{sandsaltice}.

Askaryan also predicted that at the same frequencies where the radio
Cerenkov signal is coherent, long attenuation lengths can be found in
media that occur naturally in large volumes such as ice, salt and
sand.  All past, present and proposed radio Cerenkov experiments use
one of these three media.  The first experiments to pioneer this technique
were: FORTE~\cite{forte}, a satellite antenna system that 
viewed ice over Greenland, 
GLUE~\cite{glue}, a radio dish telescope aimed at the sandy regolith on the
surface of the moon, and RICE~\cite{rice03,rice05}, an array of radio antennas 
on AMANDA strings taking data since 1999.

Most current and proposed 
radio Cerenkov experiments rely on Antarctic ice as their detection
medium.  This is due to its exceptionally large ice volume, 
the long attenuation lengths observed 
at the frequencies of interest, and the existing infrastructure and
science programs on the continent.  
Attenuation lengths in the
microwave frequencies were measured in the ice beneath the South Pole 
in 2004, and were consistent with 1~km for ice at 
-50$^{\circ}$C~\cite{southpoleice}.

\section{RICE}
\label{rice}

RICE (the Radio Ice Cherenkov Experiment) is an array of 16 antennas of 
bandwidth 200-1000~MHz
buried in the Antarctic ice beneath the Martin A. Pomerantz Observatory
(MAPO) approximately 1 km from the geographical south pole.  The antennas
are contained within a cube of ice 200~m on a side with its center
approximately 150~m below the surface.  
RICE is primarily 
searching for a radio Cerenkov signal from electromagnetic and hadronic
cascades induced by UHE neutrinos colliding with nuclei in
the ice.

The RICE detector triggers an event if at least four antennas read a 
voltage above threshold (0.1 to 1~V after amplification) within 1.25~$\mu s$.
Between 1999 and 2005, RICE's livetime was
$74.1\times10^{6}~$s and recorded 
1.035$\times10^{6}$~triggers.
No candidate neutrino events were found in offline analysis of
this data and RICE has derived an upper bound on
the flux seen in Figure~\ref{fig:neutrinos}.

RICE would also be sensitive to magnetic monopoles of intermediate
mass (IMM's) much less than the 
conventional GUT mass near $10^{17}$~GeV.  IMM's have been predicted
by Wick {\em et al.}~\cite{wick} to be relativistic
with Lorentz factors given by $\gamma=E/M_{\mathrm{IMM}}$ where
$E$ is the total energy of the monopole estimated to be $\approx10^{16}$~GeV.
Therefore, the same null
result can be used to set an upper limit on a flux of IMM's if both the
IMM energy loss in the ice surrounding RICE and the 
efficiency for triggering and reconstructing 
such events can be understood.
To model to IMM energy loss, the analysis uses 
a stochastic model for muon energy loss, replacing the IMM mass for the
muon mass and monopole charge for unit electric charge.  
From Dirac's relation for the 
magnetic monopole charge, $Z=1/(2\alpha)\approx 68$ 
where $\alpha=1/137$~\cite{dirac}.
RICE constrains the flux of intermediate mass magnetic monopoles 
to between $10^{-19}$ and $10^{-18}$/~cm$^2$/s/str  
in the range $10^8<\gamma<10^{12}$ respectively, 
and at $\gamma=10^7$ the limit is $10^{-17}$/~cm$^2$/s/str.
These are the world's best limits on
magnetic monopoles based on direct observations.

\section{ANITA}
\label{anita}

ANITA (ANtarctic Impulsive Transient Array) is an Antarctic 
balloon-borne experiment that is launched under
NASA's long duration balloon program from McMurdo station.  It consists
of an array of 32 broadband (200-1200~MHz) dual-polarization
quad-ridged horn antennas 
that view the Antarctic ice sheet
from its in-flight altitude of 37~km.  where
it is in view of 1.5$\times10^{6}$~km$^2$ of ice surface.
The first full ANITA flight was launched on December 15$^{\mathrm{th}}$, 2006,
taking 3~1/2 trips around the
continent in 35 days.  
ANITA~II has been approved and is scheduled to fly during
the 2008-2009 season.  
The author of these proceedings is a member of
the ANITA experiment.

The ANITA trigger divides the signal into frequency sub-bands 
and a coincidence requirement
provides a powerful rejection against narrow bandwidth backgrounds
and thermal noise.
A global trigger requires further coincidences
across antennas.  
While ANITA was in view
of McMurdo station,
calibration pulses were sent to the payload from above and below the 
surface.  These multi-purpose signals allowed
us to quantify our resolution on the direction of incident pulses.
Preliminary results from borehole pulses 
show an angular resolution 
of 0.2$^{\circ}$ in 
azimuth with respect to the center axis of the payload and 0.8$^{\circ}$
in zenith angle.
This reconstruction has been used to produce a  
map of the locations of signal sources.
Figure~\ref{fig:map} shows such a map produced
{\em with only 10\% of the data set} from an analysis
where the remaining 90\% of the data is blinded.  Events included in 
this plot have satisfied the requirements that the 
voltage exceed three times the thermal
noise level, angular reconstruction was successful, and
the time profile and Fourier transform are consistent with
the expectation for the radio Cerenkov signal.  All events
in this 10\% data set have been
associated with camps, travelers and automatic weather stations.  Analysis
of the remaining 90\% of the data is in progress.  ANITA expects to either
be the first to observe UHE neutrinos or set the world's
best limits in its energy range.

A test instrument with two antennas called ANITA-lite was flown in the
2003-2004 Austral summer, piggybacking on the TIGER experiment.
ANITA-lite constrained 
the UHE neutrino flux and ruled out Z-burst models, which
could have led to cosmic rays with energies that exceed the GZK cutoff 
~\cite{anitalite,zbursts}.

\section{AURA}

An experiment embedded in its detection medium, such as
RICE, can, with adequate volume, be sensitive to a 
lower energy regime than experiments where the medium and
instrumentation are well separated.  
Since the expected neutrino spectrum, 
like the measured cosmic ray spectrum,
is steeply falling, a lower threshold is a great
advantage in terms of predicted neutrino rates.

AURA (Askaryan Under ice Radio Array) is designed to
utilize existing infrastructure and technology for a radio frequency neutrino
detector at the South Pole~\cite{aura}.  The system combines RICE antennas,
electronics and control interface with the digitizer and triggering
designed for ANITA and the main board, data acquisition, boreholes and
cables used for IceCube.  
An AURA cluster consists of 
four RICE dipole antennas, a Digital
Radio Module (DRM) with electronics for
performing triggering, digitization and
transmitting data to the surface
and an Antenna Calibration Unit (ACU) for
communications and power between the surface and the DRM.  
There were three AURA clusters deployed in the 2006-2007 polar season
at the South Pole.
Single channel and cluster trigger rates have been measured 
during times when IceCube
and AMANDA were idle and also when they were taking data.  
 The combined trigger
was found to reduce noise to the level necessary for detector operation.
The AURA team plans to continue their efforts to study the radio environment
surrounding IceCube, and expand to 
a shallow UHE neutrino detector.

\section{IceRay}

Any next-generation UHE neutrino experiment aims to
move beyond the discovery stage of the field to an
era of particle physics and astrophysics measurements with 10-100~GZK
neutrinos/year.  
The proposed IceRay experiment would be 
an array of antenna 
stations deployed near the surface close to the South Pole.  
With no deep holes needed,
the cost could be kept relatively low, 
but a two-dimensional array near the
surface is only sensitive to limited depths and faces 
complications from the depth-dependent index of refraction in the
firn near the surface.  
An alternative proposal is
to expand AURA into a large three-dimensional deep array that would 
surround IceCube in boreholes dedicated for radio receivers.  
Preliminary simulations of both IceRay and AURA designs 
show that an array of either type with 18-36 stations that could
be built by 2012 could detect 4-8 neutrinos from the GZK process per year.
This would be a precursor to a larger array.  One advantage of building
a radio array at the South Pole is the possibility of observing
events in coincidence with IceCube.  The
IceRay and AURA designs are both being developed with this very desirable
feature in mind.

\section{ARIANNA}

ARIANNA is 
a proposed array of antenna stations on the surface of
the Ross Ice Shelf~\cite{arianna}.  
This detector is designed to be sensitive to not only signals
observed directly from the shower, but also signals that have been
reflected from the bottom of the ice shelf where it meets
sea water, at about 600~m depth.
This boundary is highly reflective.
Sensitivity to reflected signals means a larger range of solid angles
in the detector's acceptance compared with a surface
detector where only direct signals are observed.  Since the
ice is warmer on the ice shelf than on the continent, the
attenuation lengths are shorter, but still comparable to the depth
of the shelf.  In November 2006, 
S.~Barwick and D.~Saltzberg 
measured attenuation lengths on the Ross Ice Shelf greater
than 300~m, averaged over all depths, 
for frequencies in the range 200-1200~MHz.
A prototype station has been deployed for one year, powered by solar
panels.

\section{SalSA}
 
It has long been proposed that a neutrino detector could be deployed
in one of the large salt formations that exist in many locations around
the world~\cite{salsapaper}.  This detector concept has been termed 
Salt Sensor Array (SalSA).  
One would find 
2.5 times as many neutrino interactions per unit volume in salt compared due
ice due to its higher density.  Although the
peak power of the emitted radio Cerenkov signal is 
lower than in ice, the width of the Cerenkov cone is more 
broad~\cite{jaimemedia}.
Additionally,
an experiment in the Northern Hemisphere would view a region of the
sky not in view of an experiment at the South Pole, and could also
be more accessible.
Many salt domes exist in the Southeastern United States and 
volumes of a few~km~$\times$~few~km~$\times$~10~km are not atypical.  

Ground Penetrating Radar (GPR) experts have reported low radio
loss in salt mines in the 
US, but it is difficult to deduce attenuation length measurements
from their findings~\cite{stewartunterburger}.  Before a SalSA experiment 
can move forward,
long attenuation lengths ($>\approx250$~m) must be measured 
definitively at radio frequencies. 
Gorham {\em et al.}~\cite{salsapaper} reported attenuation lengths
$L>40$~m at 67\% confidence at the Hockley salt mine in Texas.  
Although their
mean fit values were in the region of 200~m, the uncertainties were
large, in part due to the short range of transmission, up to 45~m.
A team from UCL, UCLA and LSU that includes the author of these proceedings 
has visited the Cote Blanche salt mine in Louisiana 
and has transmitted broadband pulses 
over 300~m and from depths as much as 60~m below
the lowest mining level.  
The results from these data is forthcoming, and this topic remains
under active investigation.

\section{Conclusions}

The field of UHE neutrino detection using the radio Cerenkov technique 
has become a mature field, with existing experiments beginning to probe
the expected neutrino flux from the GZK process.   
The technique holds the capability of moving beyond the discovery stage
and into an era of making particle- and astrophysics measurements
with 10-100~UHE neutrinos per year.  I have described several proposed 
projects
that are being designed to measure
neutrinos rates at that level once operating at full scale.

\section{Acknowledgements}
The author is grateful to the following people for
providing materials and helpful comments for my talk and
these proceedings: 
Jiwoo Nam (ANITA), Bob Morris (IceRay), 
Albrecht Karle and Hagar Landsman (AURA), 
David Besson and Daniel Hogan (RICE) and Steve Barwick (ARIANNA).  
I would also like to thank
the organizers of RICH~2007 for their invitation and hospitality 
and the UCL Graduate School for their partial funding of my
travel expenses.

% The Appendices part is started with the command \appendix;
% appendix sections are then done as normal sections
% \appendix

% \section{}
% \label{}

\end{document}